\documentclass[sigconf]{acmart}
\AtBeginDocument{%
  \providecommand\BibTeX{{%
    \normalfont B\kern-0.5em{\scshape i\kern-0.25em b}\kern-0.8em\TeX}}}

\copyrightyear{2023}
\acmYear{2023}
\setcopyright{acmlicensed}\acmConference[SIGIR '23]{Proceedings of the 46th International ACM SIGIR Conference on Research and Development in Information Retrieval}{July 23--27, 2023}{Taipei, Taiwan}
\acmBooktitle{Proceedings of the 46th International ACM SIGIR Conference on Research and Development in Information Retrieval (SIGIR '23), July 23--27, 2023, Taipei, Taiwan}
\acmPrice{15.00}
\acmDOI{10.1145/3539618.3591900}
\acmISBN{978-1-4503-9408-6/23/07}

\usepackage{subfigure,graphicx}
\usepackage{booktabs}
\usepackage{multirow}
\usepackage{footnote}
\usepackage{subfigure,graphicx,array}
\usepackage{makecell}
\usepackage{bm}
\usepackage{amsmath}

\usepackage{algorithm}
\usepackage{algorithmic}
\usepackage{caption}
\usepackage{threeparttable}

\newcommand{\eg}{\textit{e.g.}}
\newcolumntype{P}[1]{>{\centering\arraybackslash}p{#1}}

\begin{document}

\title{JDsearch: A Personalized Product Search Dataset with Real Queries and Full Interactions}

\author{Jiongnan Liu}%
\orcid{0000-0002-3946-9178}
\email{liujn@ruc.edu.cn}%
\affiliation{%
\department{Gaoling School of Artificial Intelligence}%
\institution{Renmin University of China}%
\state{Beijing}%
\country{China}%
}
\additionalaffiliation{
\institution{Beijing Key Laboratory of Big Data Management and Analysis Methods}%
\state{Beijing}%
\country{China}%
}

\author{Zhicheng Dou}%
\orcid{0000-0002-9781-948X}
\email{dou@ruc.edu.cn}%
\affiliation{%
\institution{Engineering Research Center of Next-Generation Intelligent Search and Recommendation, MOE}%
\state{Beijing}%
\country{China}%
}

\author{Guoyu Tang}%
\orcid{0009-0003-9586-4652}
\email{tangguoyu@jd.com}%
\author{Sulong Xu}%
\orcid{0000-0003-0345-334X}
\email{xusulong@jd.com}%
\affiliation{%
\institution{JD.com, Inc.}
\state{Beijing}%
\country{China}%
}



\begin{abstract}

Recently, personalized product search attracts great attention and many models have been proposed. To evaluate the effectiveness of these models, previous studies mainly utilize the simulated Amazon recommendation dataset, which contains automatically generated queries and excludes cold users and tail products. We argue that evaluating with such a dataset may yield unreliable results and conclusions, and deviate from real user satisfaction. To overcome these problems, in this paper, we release a personalized product search dataset comprised of real user queries and diverse user-product interaction types (clicking, adding to cart, following, and purchasing) collected from JD.com, a popular Chinese online shopping platform. More specifically, we sample about 170,000 active users on a specific date, then record all their interacted products and issued queries in one year, without removing any tail users and products. This finally results in roughly 12,000,000 products, 9,400,000 real searches, and 26,000,000 user-product interactions. We study the characteristics of this dataset from various perspectives and evaluate representative personalization models to verify its feasibility. The dataset can be publicly accessed at Github: \textbf{\url{https://github.com/rucliujn/JDsearch}}.


\end{abstract}






\maketitle
\section{Introduction}
\label{sec:intro}

In search engines, the queries issued by users are mostly broad and vague~\cite{amb_query, amb_query2}. This problem is extremely crucial for product search scenarios, since users may only issue the query such as ``Microsoft'' (the brand), ``Personal Computer''(the category), or ``Surface''(part of the product name), instead of the exact name of the product ``Microsoft Surface Pro 9'' to find and purchase it. If online shopping platforms only use ad-hoc ranking models to provide results, the products that users intend to purchase may be displayed at the bottom positions, which makes users unsatisfied and the sales of platforms low. To solve this problem, similar to the personalized web search area, many personalized product search approaches~\cite{HEM,ALSTP,ZAM,DREM,DREM-HGN,TEM,RTM,CAMI} have been proposed. These methods utilize user histories to rank products considering both the relevance to the current query and user interests. Applying this strategy, online platforms can show products that users want to buy at top positions.

Unfortunately, the lack of large-scale datasets based on real user behaviors blocks the studies of personalized product search. The most widely-used personalized product search datasets are simulated Amazon sub-category datasets~\footnote{\url{http://jmcauley.ucsd.edu/data/amazon/}}. These datasets are collected from the interactions between products and users in the amazon online shopping websites from 1996 to 2014. However, these datasets are originally developed for evaluating recommendation models and only contain the reviews of users and products, \textbf{without the genuine queries issued by users}. To apply these datasets in personalized product search experiments, researchers usually concatenate the terms in the category lists of products to build pseudo queries~\cite{LSE,HEM}. These generated queries can somehow resemble the ambiguity of actual queries, but in reality, users may issue diverse queries based on several fields of the products they want to buy, including brands, categories, names, and detailed descriptions. This may make the experimental performances based on pseudo queries in these datasets differ from the real product search scenarios. Furthermore, these datasets also put some restrictions on the products and users. For example, they \textbf{only include products and users whose total amounts of reviews are larger than five (5-core products and users)}, which is an obstacle for few-shot model studies. Another drawback of this dataset is that each dataset \textbf{solely contains products belonging to the same category}. However, user histories usually contain products belonging to a variety of categories in reality, and modeling user interests across diverse categories may be beneficial for personalization. As a result, the simulated user behaviors in these datasets can vary from the patterns in practice, which makes the results on them unreliable. Thus, the models designed using these datasets may be not applicable in real situations. There also exists a few open-resource datasets such as DIGINETICA~\footnote{\url{https://competitions.codalab.org/competitions/11161}}. However, these datasets are in a small scale and few previous approaches utilize them. To conclude, research progress in the personalized product search area requires a large-scale dataset containing real user behaviors, but existing datasets are unsatisfactory.

In this paper, we construct and release a new dataset, \textbf{JDsearch}, based on a popular Chinese online shopping platform JD.com. JDsearch dataset is a ready-to-use and well-documented anonymized dataset. It is licensed under CC BY-NC-SA 4.0~\footnote{\url{https://creativecommons.org/licenses/by-nc-sa/4.0/}}. The dataset contains about 170,000 users, 12,000,000 products, 9,400,000 real searches, and 26,000,000 user-product interactions between them. During the construction of this dataset, we record real user queries and reserve all products belonging to all categories, regardless of their popularity. Users with various history lengths are also included in this dataset. To protect the privacy of users, we anonymize all the sensitive information including ids and texts in JDsearch dataset.  We analyze this dataset from several perspectives including users, products, queries, and personalization potential to show its advantages. Finally, we test a wide range of existing personalized product search models on this dataset to show the feasibility of using it to conduct personalization studies. Overall, JDsearch dataset has the following advantages and can inspire some new research topics :

1) Different from the pseudo queries in the Amazon dataset, the queries in JDsearch dataset are real. This can make the experimental results on this dataset closer to the online serving scenarios.

2) We reserve all the products belonging to diverse categories in JDsearch dataset, which can support future studies considering multiple user preferences and few-shot products.

3) We include both users with extremely long and short histories, which may lead the future approaches to design different strategies for different kinds of users.

\section{Related Work}
\label{sec:related_work}

Previous personalized product search works usually utilize the Amazon sub-category datasets to conduct experiments. These datasets are originally recommendation datasets, which contain product reviews and metadata on the Amazon websites from May 1996 to July 2014. As the scale of the overall dataset is too huge for models to personalize results, researchers usually select some sub-category datasets to set up experiments and these datasets only contain products belonging to one single category. Besides, these datasets also filter out products and users whose reviews are less than five to obtain the denser 5-core datasets. Since these datasets only contain the review information of products, researchers need to heuristically build pseudo queries from the metadata of products. Previous approaches commonly utilize the categories of products to build queries: they concatenate the words in the category lists and remove the duplicated words and stopwords to construct the queries for corresponding products. In this way, the generated queries can mimic the ambiguity of real queries. However, users don't issue queries solely based on the category information of the products they want to purchase in reality. 

Except for the widely-used simulated Amazon sub-category products, there only exists a few open-resource datasets. DIGINETICA or CIKMCUP2016 dataset is only utilized in two previous works~\cite{IHGNN,GraphSRRL}. This dataset is collected from a Russian online shopping platform. However, over half of the purchase behaviors in this dataset are done by guests~\cite{IHGNN}. In other words, these behaviors don't have user ids and cannot be utilized for personalization. After filtering these anonymous behaviors, the remaining dataset is quite small and difficult for conducting personalization experiments. Besides the DIGINETICA dataset, there also exist some online competitions containing users' online shopping histories. However, to the best of our knowledge, no previous studies apply these competition datasets in their experiments.

\section{JDsearch: A New Personalized Product Search Dataset}

As we mentioned in Section~\ref{sec:intro}, queries in Amazon datasets are heuristically constructed by the categories of products~\cite{LSE}. In contrast to this hypothetical situation, in reality, users can issue a variety of queries based on different attributes such as titles, brands, and categories to search products. Thus, the simple pseudo query generation process in Amazon datasets cannot mimic the real search behaviors well. 
Further, Amazon sub-category datasets also put several limitations on the products and users. First, it only contains products and users that have at least 5 reviews or purchases (5-core products and users) and leaves others out. Second, in each sub-category dataset, products in the corpus all fall under the same category.  However, in real situations, what personalization models face is that tail users and products are common while products in user histories may belong to diverse categories. In conclusion, the above restrictions may cause the \textbf{simulated Amazon datasets to differ from the real situation}. This may lead the methods developed, tested, and assessed using Amazon datasets to become inapplicable in reality. As a result, it is essential to construct and release a new dataset based on real user behaviors.

\subsection{Dataset Construction}

To solve the problems in Amazon datasets and support research in the personalized product search area, we construct a new personalized product search dataset based on genuine user behaviors from a Chinese online shopping platform JD.com. The preprocessing pipeline for constructing JDsearch dataset is as follows:

\textbf{1) User Behavior Collection.} First, we randomly sample about 170,000 users who have issued queries on the platform on a specific date, 2022-10-17. The histories of users are formed by their \textbf{issued queries} and \textbf{interacted products} from 2021-10-18 to 2022-10-17. Unlike the Amazon datasets, we place no restrictions on the categories or populations of the history products or the history lengths of users. Therefore, all the products in the dataset corpus belong to a variety of categories, which makes the user histories more diverse. In particular, we include all types of user behaviors including click, add to cart, follow and purchase, and record these \textbf{interaction type labels} in JDsearch dataset, which can provide signals for future works considering multiple feedback. The \textbf{timestamps} of these actions are also recorded in our dataset which can offer temporal information to models. Specifically, we don't require the \textbf{interactions with products must be under issued queries}, which means users may interact with products from diverse channels including search, recommendation, and casual browsing. For example, if a user purchases an iPhone by searching "smart cellphone" and then clicks the AirPods product through the recommendation systems, these behaviors will all be recorded in JDsearch but Amazon datasets may only record the first one. In summary, we record the historical queries, interacted products, interaction type labels, and their corresponding timestamps of users in our JDsearch dataset.
:
\textbf{2) Product Metadata Collection.} Then, for the product meta information, we record the \textbf{names, categories, brands, and shops of products} in our dataset. There exist \textbf{four-level categories} in the JD online shopping platform and we reserve all of them. However, the JDsearch dataset doesn't contain the related item relationships (such as "bought together" in the metadata part of Amazon datasets) because the platform doesn't release these data. 

\textbf{3) Anonymization.} Next, since this dataset is collected from a commercial shopping platform, we need to anonymize JDsearch to remove personal private information. For the ids of products, brands, categories, and shops, we randomly hash them to numbers in a wide range. For the textual information, we first conduct word segmentations for texts including queries, product titles, category names, etc. Then, we randomly hash these term ids to integers, too. 

\textbf{4) Dataset Partition.} Finally, similar to the popular leave-one-out evaluation methods in recommendation systems, we use the last queries of the users issued on 2022-10-17 as the \textbf{test queries}. The behaviors before the last queries can be used to train models. Different from the Amazon datasets that don't have the displayed results of queries, we obtain the exposed product lists and their labels under the test queries issued by users in this commercial platform. We further remove the duplicated products in these displayed results and reserve at most 200 products as the \textbf{candidate product lists} for these test queries. Besides, users who don't interact with any products under their last queries (about 2,000 users) are removed from the test part of JDsearch dataset. Because of including the exposed products, personalization models can conduct fine-grained ranking in JDsearch dataset instead of the coarse-grained retrieving in Amazon datasets. In a nutshell, we record the test queries, their candidate product lists, and the labels for candidates for each user in JDsearch dataset.

The fields and explanation in user behavior data and product meta data in JDsearch dataset are shown in Table~\ref{tlb:field}.  The detailed content and format description of our JDsearch dataset can be found in our repository.~\footnote{\url{https://github.com/rucliujn/JDsearch}}. 

\begin{table}[!tbp]
 \setlength{\abovecaptionskip}{0.1cm}
  \setlength{\belowcaptionskip}{0.1cm}
\centering
    \caption{Field and explanation in the datasets}\label{tlb:field}
    \begin{tabular}{p{.37\linewidth}@{}|p{.63\linewidth}@{}}
        \toprule
         Filed &  Explanation  \\
         \hline
         \multicolumn{2}{l}{\emph{User Behavior Data}} \\
         \hline
         query & the anonymized term ids of the test query. \\
         candidate\_wid\_list & the anonymized id list of candidate products displayed under the test query. \\
         candidate\_label\_list & the corresponding label for the candidate products. \\
         history\_qry\_list & the sequence of anonymized term ids of issued queries in user histories.\\
         history\_wid\_list & the sequence of anonymized ids of products in user histories. \\
         history\_type\_list & the sequence of interaction levels in user histories. \\
         history\_time\_list & the sequence of timestamps of interactions in user histories. \\
         \hline
         \multicolumn{2}{l}{\emph{Product Meta Data}} \\
         \hline
         wid & the anonymized id of the product. \\
         name & the anonymized term ids of the product's name. \\
         brand\_id & the anonymized id of the product's brand. \\
         brand\_name & the anonymized term ids of the product's brand name. \\
         category\_id\_\{1,2,3,4\} & the anonymized ids of the the product's four level categories. \\
         category\_name\_\{1,2,3,4\} & the anonymized term ids of the product's four level categories' names. \\
         shop\_id & the anonymized id of the product's shop. \\
        \bottomrule
    \end{tabular}
\end{table}

Overall, compared with the Amazon sub-category datasets, The main advantages of our JDsearch dataset are: \textbf{1) it includes real user queries}, \textbf{2) it reserves all products with different categories and comprises both cold and popular products}, \textbf{3) it contains various types of users whose history lengths are diverse}.  Furthermore, we also record all interactions with various types and prepare the candidate product lists for test queries. We summarize the characteristics of JDsearch dataset and the previous sub-category dataset in Table~\ref{tlb:char}. In the following part, we will analyze our dataset to demonstrate these advantages.

\begin{table}[!tbp]
 \setlength{\abovecaptionskip}{0.1cm}
  \setlength{\belowcaptionskip}{0.1cm}
\centering
    \caption{characteristics of the datasets}\label{tlb:char}
    \begin{tabular}{p{.30\linewidth}@{}p{.30\linewidth}<{\centering}p{.30\linewidth}<{\centering}}
        \toprule
         Characteristic & Amazon datasets &  JDsearch dataset \\
         \midrule
         Query & Artificial & Real  \\
         Item popularity & {5-core items} & {All items}\\
         Item category & {Same category} & {Diverse categories} \\
         User popularity & 5-core users & All users  \\
         \multirow{2}{*}{Interaction type} & \multirow{2}{*}{Purchase}   & {Click, add to cart, follow, purchase} \\
        \bottomrule
    \end{tabular}
\end{table}

\section{Dataset Analysis}
First, we provide basic statistics in our dataset and some Amazon sub-category datasets in Table~\ref{tlb:sta}. Compared with widely used Amazon sub-category datasets, we don't filter out the products which have only been interacted with a few times and keep track of all the interactions in user history. Thus, we can find that JDsearch dataset is much sparser and the average user history length is much longer, making it more challenging for models to capture user interests and conduct personalized ranking. Besides, our dataset contains more test queries than Amazon datasets, which can make the evaluation more stable and convincing. 

\label{sec:dataset}
\begin{table*}[!tbp]
\begin{threeparttable}
 \setlength{\abovecaptionskip}{0.1cm}
  \setlength{\belowcaptionskip}{0.1cm}
\centering
    \caption{Statistics of the datasets}\label{tlb:sta}
    \begin{tabular}{lccccc}
        \toprule
         Dataset & \emph{Cell Phones \& Accessories} & \emph{Clothing, Shoes \& Jewelry} & {\emph{Sports \& Outdoors}} & \emph{Electronics} &  JDsearch \\
         \midrule
         \#Users & 27,879 & 39,387 & 35,598 & 192,403 & 173,831 \\
         \#Items & 10,429 & 23,033 & 18,357 & 63,001 &  12,872,636 \\
         \#Interactions & 194,439 & 63,001 & 278,677 & 1,689,188 & 26,667,260 \\
         \#Test Queries\tnote{*} & 426 & 6,939 & 472 & 3,221 &  171,728 \\
         \bottomrule
    \end{tabular}
    \begin{tablenotes}
        \footnotesize
        \item[*] The numbers of test queries in Amazon datasets are calculated based on the sequential-based dataset division in previous works~\cite{TEM}.
      \end{tablenotes}
\end{threeparttable}
\end{table*}

\subsection{Product and User Analysis}

In this part, we investigate the characteristics of products and users in JDsearch dataset.

\subsubsection{Product's Interaction Frequency Analysis} 
\begin{figure}[!tbp]
 \setlength{\abovecaptionskip}{0.1cm}
 \setlength{\belowcaptionskip}{0.1cm}
  \centering
  \includegraphics[width=0.8\linewidth]{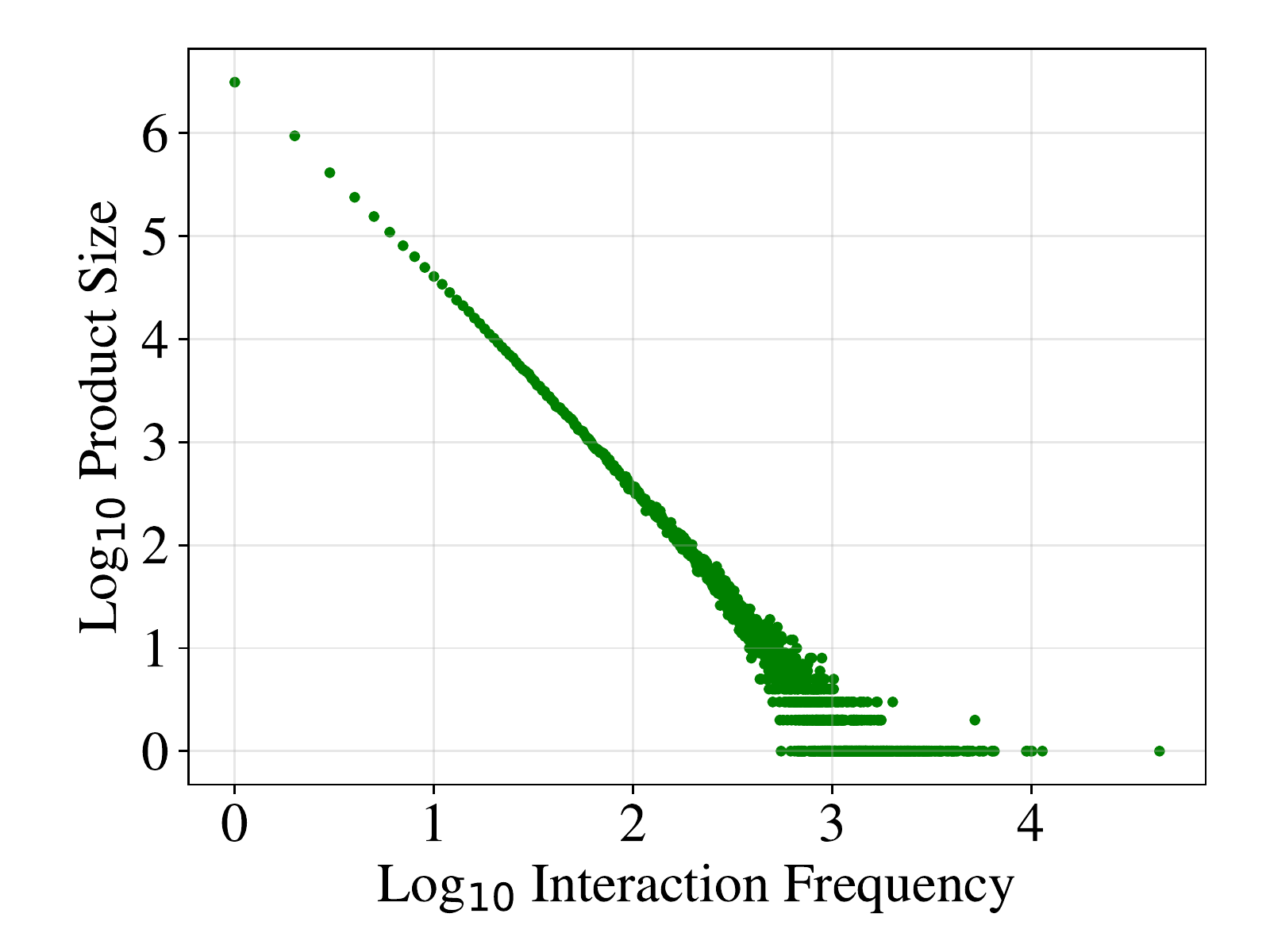}
  \caption{The log-log distribution of product's interaction frequency}
  \Description{The log-log distribution of product's interaction frequency}
  \label{fig:prod_dist}
\end{figure}

First, we demonstrate the distribution of the product's interaction frequency in the JDsearch dataset in Figure~\ref{fig:prod_dist}. We can find \textbf{the distribution of product's interaction frequency aligns with power law distribution}, which means that many products have only interacted with users a few times (cold products) while the amounts of hot products (frequently interacted with users) are relatively limited. This phenomenon suggests that the preprocess manipulation of only including 5-core products in Amazon datasets may destruct the continuous user behaviors in reality and may result in different performances compared with retaining all products. Besides, this configuration may help models filter out noisy histories and makes personalization easier. However, in reality, personalization models need to discriminate which part of user histories is more important for personalization and more related to current queries.

\subsubsection{User History Analysis} 
\begin{figure}[!tbp]
 \setlength{\abovecaptionskip}{0.1cm}
 \setlength{\belowcaptionskip}{0.1cm}
  \centering
  \includegraphics[width=0.8\linewidth]{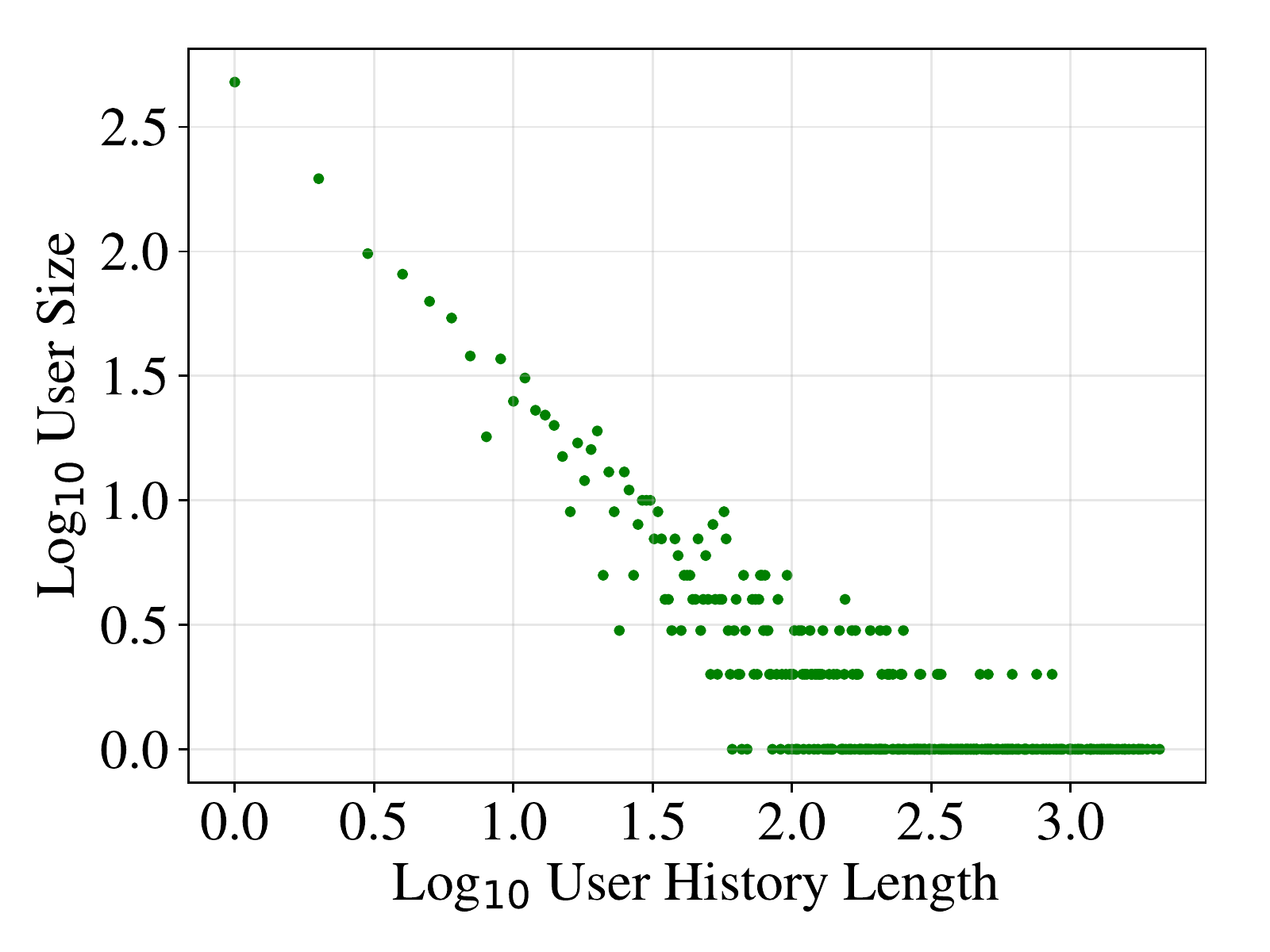}
  \caption{The log-log distribution of user history length}
  \Description{The log-log distribution of user history length}
  \label{fig:user_dist}
\end{figure}

Then, we analyze the pattern of user histories. The frequencies of user history length in JDsearch dataset are shown in Figure~\ref{fig:user_dist}. We can discover that there exist users with very short histories (only have one or two interacted products). Meanwhile, some users that have extremely rich interaction histories are also included in our dataset. This phenomenon indicates that \textbf{users may have different personalities: some users rarely do online shopping while some users frequently browse products online.} Thus, it may stimulate fresh research on designing different personalization strategies for users with different characters. For example, models may incorporate more universal preferences from all users into the user modeling while the users have limited histories and pay more attention to personal interests while they have rich interactions.

\begin{figure}[!tbp]
 \setlength{\abovecaptionskip}{0.1cm}
 \setlength{\belowcaptionskip}{0.1cm}
  \centering
  \includegraphics[width=0.8\linewidth]{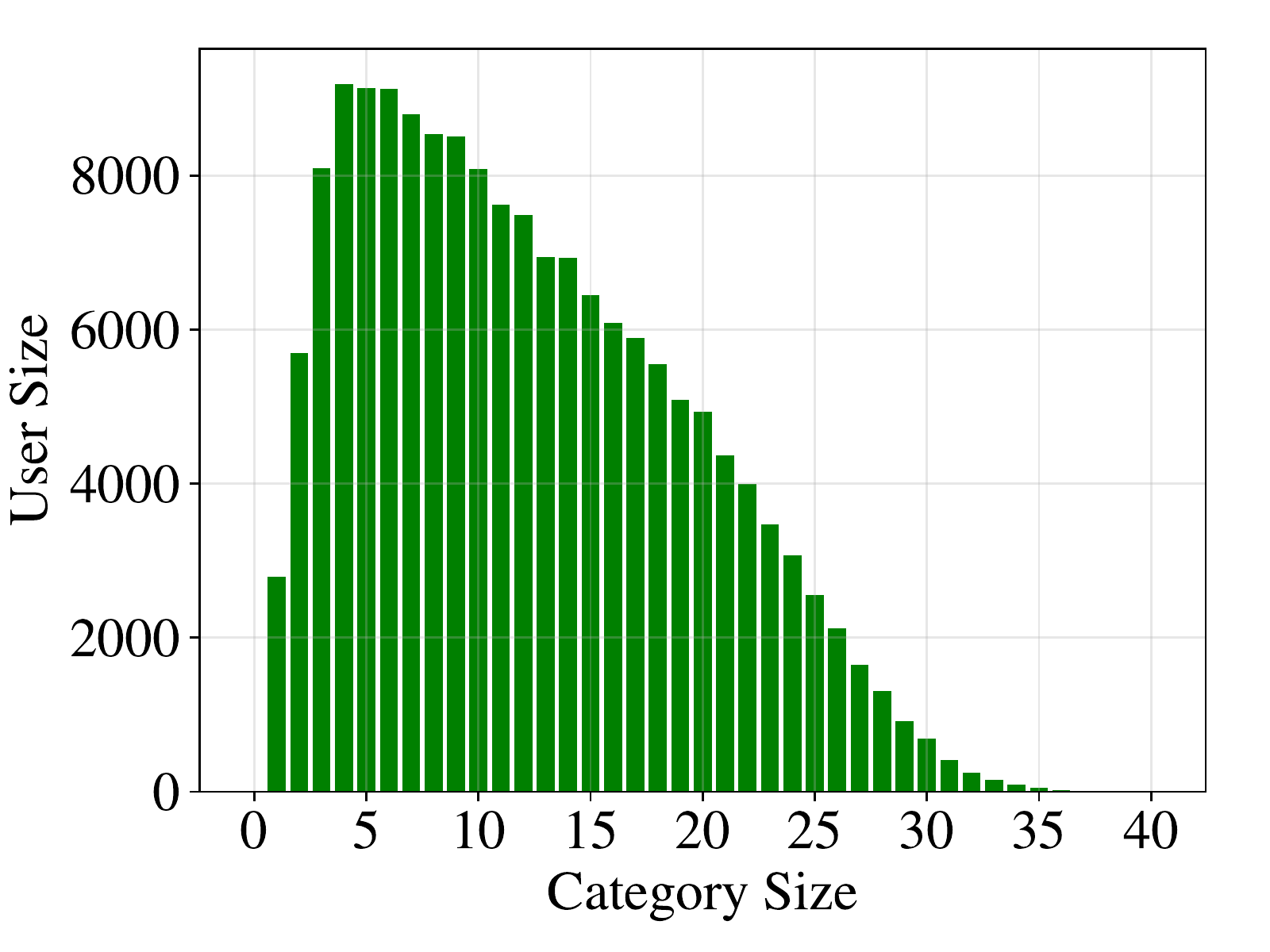}
  \caption{The category sizes of products in user histories}
  \Description{The category sizes of products in user histories}
  \label{fig:cate_dist}
\end{figure}

Besides, we also show the numbers of products' first-level categories in user histories in Figure~\ref{fig:cate_dist}. For a fair comparison, we only investigate users whose history length is larger than five in our dataset, since Amazon datasets only include 5-core users. We can find that \textbf{most users' histories are formed with products belonging to different categories.} Different from this characteristic, in the Amazon 5-core sub-category datasets, user histories only contain products belonging to one certain top-level category. However, customers' interests in one category may extend to other categories, hence only containing products belonging to one category can be harmful to personalization. Therefore, it may be better to include products with all categories in user histories as we did in the JDsearch dataset.


\subsection{Query Analysis}
\begin{figure}[!tbp]
 \setlength{\abovecaptionskip}{0.1cm}
 \setlength{\belowcaptionskip}{0.1cm}
  \centering
  \includegraphics[width=0.8\linewidth]{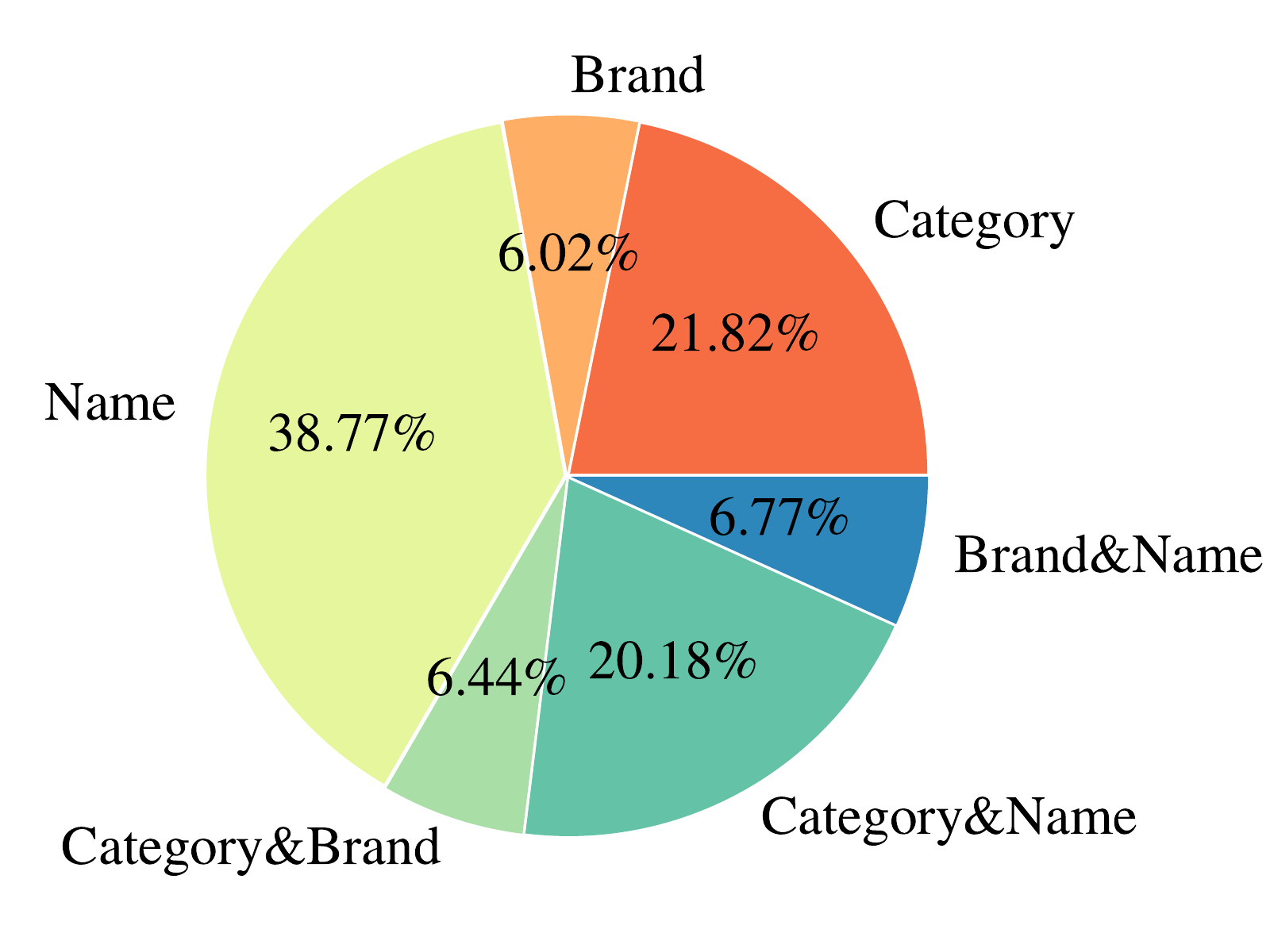}
  \caption{Distribution of query types}
  \Description{Distribution of query types}
  \label{fig:query_dist}
\end{figure}

As we mentioned in Section~\ref{sec:intro}, the most crucial problem in Amazon datasets is that the queries in them are pseudo ones and are generated by the categories of products. These queries can somehow reflect the ambiguity of the real issued queries by users. However, based on our JDsearch dataset gathered from real user behaviors, we observe that \textbf{the source of query terms can be the brands, names, or even detailed descriptions of products, not only the categories of the products they want to purchase}.  Specifically, depending on the sources of the  terms, we categorize queries in JDsearch dataset into the following types: 
\begin{itemize}
    \item \textbf{Category}: All the query terms belong to the interacted product's category (\eg, cellphone).
    \item \textbf{Brand}: All the query terms belong to the interacted product's brand (\eg, Apple).
    \item \textbf{Name}: All the query terms belong to the interacted product's name (\eg, Surface Pro 8).
    \item \textbf{Category\&Brand}: All the query terms belong to the interacted product's category and brand (\eg, Apple cellphone).
    \item \textbf{Category\&Name}: All the query terms belong to the interacted product's category and name (\eg, Laptop Surface).
    \item \textbf{Brand\&Name}: All the query terms belong to the interacted product's brand and name (\eg, Microsoft Surface).
\end{itemize}
We show the type distribution of queries in Figure~\ref{fig:query_dist}. We can observe that the terms of most queries issued by users only derive from the names of products. After that, users also usually enter queries based on the categories or the combination of categories and names.  Besides, there also exist queries based on the brands of products. Noticing that in this part, we only investigate queries that can be categorized into the above types. There still exist many queries whose terms originate from other fields of products. These findings show that the naive strategy of constructing queries from categories of products is inadequate, \textbf{a better way may be to utilize diverse fields of products to assemble pseudo queries}. 

\subsection{Potential of Personalization}

In this part, we analyze the personalization potential in our JDsearch dataset.

\subsubsection{Query Ambiguity Analysis} First, we analyze the test queries and their candidates in our dataset. We provide how many test queries have candidate products that come from various first-level categories (category ambiguous)  and various brands (brand ambiguous). Among all 171,728 test queries, 114,955 are category ambiguous, 109,874 are brand ambiguous, and 68,478 are both category and brand ambiguous. Through these statistics, we can find that \textbf{most test queries in our JDsearch dataset are vague}. In this case, it is hard for simple ad-hoc ranking methods to provide satisfying results for users, so personalization is required. 

Then, we analyze all the history queries and show their ambiguity. To quantify the ambiguity of queries, we propose a metric interaction entropy(IE) for queries. The interaction entropy is a natural extension of click entropy~\cite{pclick} and can be calculated from the corresponding interactions under queries. Generally speaking, when the interaction entropy of a query is larger than or equal to one, we can infer that this query is an informational query and may be ambiguous. The interaction entropy is calculated as follows:
\begin{align}
    {\mathrm{IE}}(q) &= \sum_{p \in \mathcal{I}(q)} - P(p|q) \log_2 P(p|q) \\
    P(p|q) &= \frac{|{\mathrm{Interaction}}(q,p)|} {\sum_{p' \in \mathcal{I}(q)} |{\mathrm{Interaction}}(q,p')|},
\end{align}
where $\mathcal{I}(q)$ is the collection of products interacted (including clicked, added to cart, followed, and purchased) with users under query $q$, $P(p|q)$ is the percentage of interactions on product $p$ among all interactions under query $q$. we calculate the interaction entropy for all repeated queries in JDsearch dataset and show the numbers of queries whose IE is less than / equal to / larger than one in Table~\ref{tlb:entropy}. From the statistics, we can find that \textbf{most queries in our corpus are ambiguous and different products have been interacted under them}. This can also verify the personalization is necessary in our dataset.
\begin{table}[!tbp]
 \setlength{\abovecaptionskip}{0.1cm}
  \setlength{\belowcaptionskip}{0.1cm}
\centering
    \caption{Interaction entropy distribution of repeated queries in the JDsearch dataset}\label{tlb:entropy}
    \begin{tabular}{lccc}
        \toprule
         Entropy & IE $< 1.0$ & IE $= 1.0 $ & IE $> 1.0$\\
        \midrule
         \#Queries & 147,095 (18.69\%) & 231,638 (29.44\%) & 408,195(51.87\%) \\
         \bottomrule
    \end{tabular}
\end{table}

\subsubsection{User Interest Analysis} 
\begin{figure}[!tbp]
 \setlength{\abovecaptionskip}{0.1cm}
 \setlength{\belowcaptionskip}{0.1cm}
  \centering
  \includegraphics[width=\linewidth]{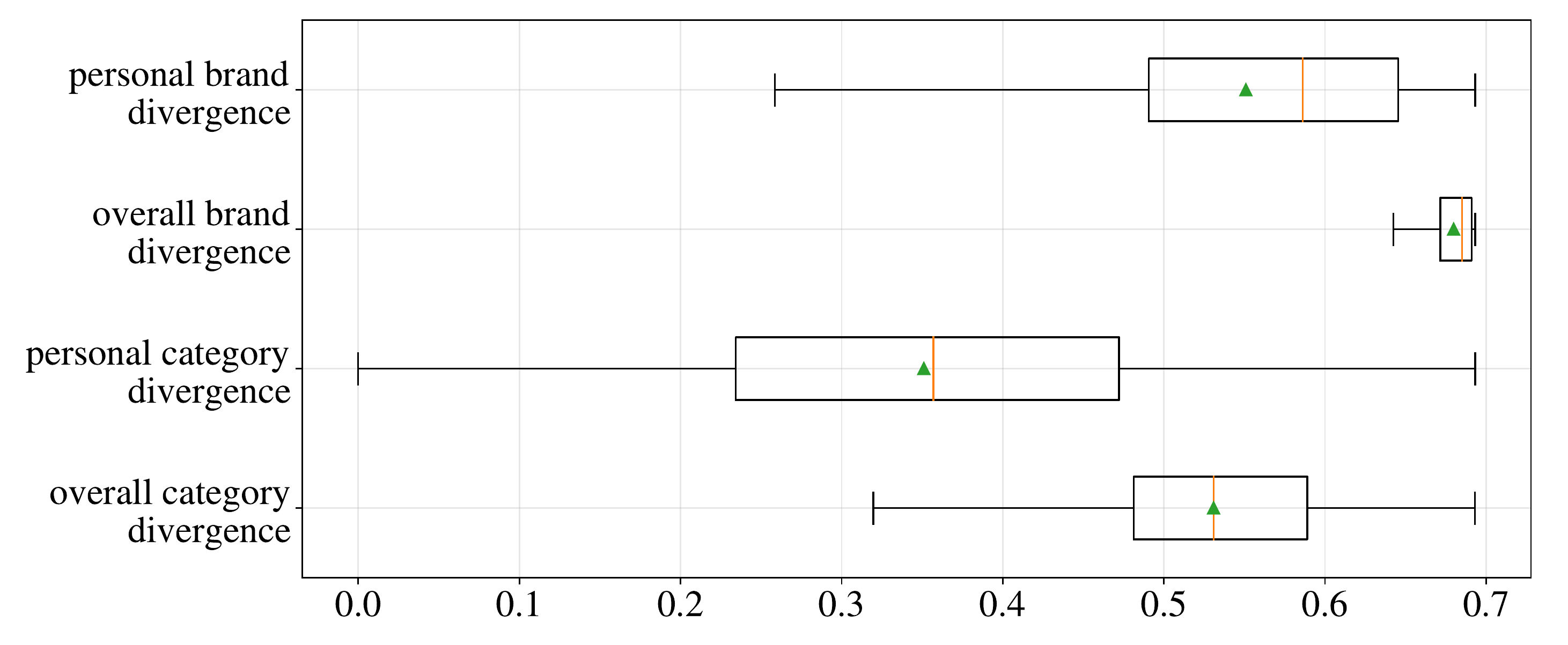}
  \caption{The boxplot of the personal and overall interest divergence of each user}
  \Description{The boxplot of the personal and overall interest divergence of each user}
  \label{fig:divergence}
\end{figure}

Finally, we investigate the user interest distribution in JDsearch dataset. We define user interest distribution as the top-level category or brand distribution of their historically interacted products. If the JDsearch dataset supports personalization, the user's interests should be consistent, which means the interest distribution of users in their early histories and late histories should be similar. To verify this assumption, we calculate three interest distributions and two distribution divergences: valid distribution, support distribution, overall distribution, personal divergence, and overall divergence. The \textbf{valid distribution} of one certain user $u$ is calculated based on her own late histories (the latest ten interactions). The \textbf{support distribution} of $u$ is calculated based on her own early histories (interactions before the latest ten ones). The \textbf{overall distribution} is calculated based on all users' early histories. The \textbf{personal divergence} of $u$ is calculated between her valid distribution and her support distribution. The \textbf{overall divergence} is calculated between her valid distribution and the overall distribution. All these divergences are calculated by JS divergence. We calculate the personal divergence and overall divergence of each user $u$ based on the first-level category and brand respectively. We argue that if user interests are continuous and personalization is rational, the personal divergence should be smaller than the overall divergence. We show the results in Figure~\ref{fig:divergence}.

In the boxplot figure, since there must exist some users' late purchase behaviors are totally different from the early ones and the overall preferences, the upper bound of the personal divergences and overall divergences are all close to $\log 2$, which is the upper bounds of JS divergence. However, we can infer that most users' personal divergence is smaller than the overall divergence, which suggests that \textbf{most users' interests in JDsearch dataset are in consistency} and personalization can be effective in our dataset.

\section{Experiments and Analysis}
\label{sec:settings}
The JDsearch dataset can be used for personalized product search studies to conduct experiments. In this part, we evaluate representative personalized product search models in our JDsearch dataset to verify the feasibility of performing personalization in this dataset. Further, we also conduct experiments on several dataset variations to show the different characteristics of our dataset compared with Amazon datasets.

\subsection{Settings and Evaluation Metrics}
In our experiments, we train the models using the behaviors with queries in user histories. For inference, as we mentioned in Section~\ref{sec:dataset}, we use the last queries of users as the test queries. So we apply the trained models to rank the candidate product lists to evaluate their performances. For evaluation metrics, we use MRR@200, Precision@1, and NDCG@10 to evaluate the ranking results.

\subsection{Benchmark models}

We experiment with the following ad-hoc and personalized models:

\textbf{BM25}: BM25~\cite{bm25} is a classical sparse ad-hoc retrieval model.

\textbf{QEM}: QEM only considers the matching scores between products and queries and can be regarded as a neural ad-hoc model.

\textbf{HEM}: HEM~\cite{HEM} is a latent vector based personalized model. It builds the representations of users and items by generative language models based on reviews.

\textbf{DREM}: DREM~\cite{DREM} is a KG based personalized model. It utilizes the metadata of items to establish a knowledge graph.

\textbf{AEM, ZAM}: AEM~\cite{ZAM} is an attention-based personalized model. It aggregates the user historical interacted items with the current query to construct a query-specific user profile. ZAM improves AEM by concatenating a zero vector to the item list to adjust the extent of conducting personalization.

\textbf{TEM}: TEM~\cite{TEM} is a transformer-based personalized model. It upgrades the attention layer in AEM with transformer encoder.

\textbf{HGN}: HGN~\cite{DREM-HGN} integrates \textbf{DREM} and \textbf{ZAM} models. It uses the relations in KG to boost the representations of products and users. 

\subsection{Implementation Details}
As we anonymize the textual information in the dataset, which is much sparser than Amazon datasets, it is hard to optimize the word embeddings from scratch. So we adopt word2vec~\cite{word2vec} to initialize the word embedding in models. Besides, different from the settings in Amazon datasets, products in the test set may have not been seen in the training process (they may be cold products). In addition to the large scale of product corpus, simply optimizing the product embedding table as previous models may cause extreme overfitting. Therefore, we obtain products' representations in models by calculating dynamic vectors based on their texts including title, brand, and category. We use the average term vectors of products' texts and apply a simple non-linear function to obtain product representations. Due to the privacy protection policy, we don't have the review information of users and products. Thus, we remove the generative language modeling loss for items and users from all models. Thus, in HEM~\cite{HEM} and DREM~\cite{DREM}, again, we apply the non-linear function to build user embeddings from all the terms in their history interacted items. For the KG-enhanced models, as JD commercial platform doesn't provide the related product relationships (such as ``bought together'' and ``also bought') to us, so we only utilize the category and brand information to build the item-attribute graph. For the hyper-parameter settings, we set the maximum user history length in all models as 50. We set the embedding dimension as 128 and train them for 30 epochs. The number of transformer layers in transformer-based models is chosen from \{1, 2\}. The number of attention heads in attention-based models is set as 8.  For all negative sampling in models, we randomly choose five negative samples in uniform distribution over the corpus.

\subsection{Overall Results}
\begin{table}[t]
 \center
 \setlength{\abovecaptionskip}{0.1cm}
 \setlength{\belowcaptionskip}{0.1cm}
 \caption{Overall performances of models. The best and the second results are denoted in bold and underlined fonts respectively.}\label{tlb:overall}
  \begin{tabular}{p{.18\linewidth}<{\centering}|p{.10\linewidth}|p{.15\linewidth}<{\centering}p{.15\linewidth}<{\centering}p{.15\linewidth}<{\centering}}
  	\toprule
  	 \multicolumn{2}{c|}{Model}& MRR & Prec & NDCG \\ 
  	 \midrule 
 \multirow{2}*{\shortstack{Ad-hoc}} 
  & BM25 & 0.1114 & 0.0402 & 0.0940 \\
  & QEM & 0.1774 & 0.0728 & 0.1705  \\
  	 \midrule 
 \multirow{6}*{\shortstack{Personalized}} 
  & HEM & 0.1955 & 0.0847 & 0.1905  \\
  & DREM & 0.1647 & 0.0632 & 0.1578  \\  
  & HGN &  0.1662 & 0.0634 & 0.1591 \\
  & AEM & \underline{0.1971} & \underline{0.0851} & \underline{0.1920} \\
  & ZAM & 0.1969 & 0.0849 & \underline{0.1920} \\
  & TEM & \textbf{0.2229} & \textbf{0.1049} & \textbf{0.2192}  \\
	\bottomrule
  \end{tabular}
\end{table}

The overall results are shown in Table~\ref{tlb:overall}. We can find that all the neural product search models overperform the sparse retrieval method BM25, which shows the effectiveness of neural ranking models. Except for the KG-enhanced DREM and HGN models, all the personalized product search models achieve improvements over the ad-hoc QEM models, demonstrating the feasibility of conducting personalization in our dataset. The poor performances of KG-based models may result from that we don't have efficient meta relationships among products and attributes. A promising way of solving this issue is to select some user histories to build a denser knowledge graph. This experiment proves that it is possible to train and evaluate personalized product search models in JDsearch dataset. However, the improvements over the ad-hoc search models are relatively incremental, which indicates that there still exists potential research topics for more accurate personalization using our JDsearch dataset.

\subsection{Dataset Ablation Study}
As we introduced in Section~\ref{sec:intro}, some manipulations in previous Amazon datasets are obstacles for personalized product search research. For example, the pseudo queries generated artificially may be different from the real queries issued by users. To further show the characteristics and advantages of our JDsearch dataset, we resemble some operations in processing Amazon datasets and adjust the construction procedure of JDsearch dataset to produce several dataset variants. These variants include :

\textbf{JDsearch$\rm{_{fakequery}}$}: We replace all the user historical queries with the artificial queries generated by concatenating the corresponding interacted products' category terms but keep the real  test queries. 

\textbf{JDsearch$\rm{_{samecate}}$}: We only reserve the products and their corresponding queries (if they have any) in user histories whose first-level category is the same as the products that the users finally interacted with under the test queries.

\textbf{JDsearch$\rm{_{w/o\;cold}}$}: We only reserve the products and their corresponding queries (if they have any) in user histories that have at least five interactions with users.

\textbf{JDsearch$\rm{_{short}}$}: We don't change the training process of models but only keep users whose history length is not larger than two in the test part.

\textbf{JDsearch$\rm{_{long}}$}: We don't change the training process of models but only keep users whose history length is larger than three hundred in the test part.

Then we train and evaluate the TEM model in these dataset variants and the results are shown in Table~\ref{tlb:abl}.

\begin{table}[t]
 \center
 \caption{The performances of TEM models in different dataset variants.}\label{tlb:abl}
  \begin{tabular}{p{.35\linewidth} p{.12\linewidth}<{\centering} p{.12\linewidth}<{\centering} p{.12\linewidth}<{\centering}}
  	\toprule
  	 {Dataset}& MRR & Prec & NDCG \\ 
  	 \midrule
            JDsearch & 0.2229 & 0.1049 & 0.2192  \\
            \quad $\rm{JDsearch_{fakequery}}$ & 0.1644 & 0.0661 & 0.1539 \\
            \quad $\rm{JDsearch_{samecate}}$ & 0.1925 & 0.0840 & 0.1869 \\
            \quad $\rm{JDsearch_{w/o\;cold}}$ & 0.2296 & 0.1088 & 0.2272\\
            \quad $\rm{JDsearch_{short}}$ & 0.2080 & 0.1012 & 0.1979 \\
            \quad $\rm{JDsearch_{long}}$ & 0.2294 & 0.1096 & 0.2215\\
	\bottomrule
  \end{tabular}
\end{table}

We can find that all these performances on different dataset variants are significantly different. These results show that the manipulations in Amazon sub-category datasets have effects on the model's performances and may skew the evaluation of approaches. We can find by replacing the queries in the training set with the pseudo queries in the \textbf{JDsearch$\rm{_{fakequery}}$} dataset, the model's performance while facing real queries issued by users can be poor. Only including products belonging to the same category also harms the user interest construction as evidenced by the results on \textbf{JDsearch$\rm{_{samecate}}$}. The operations of removing the cold products from user histories in \textbf{JDsearch$\rm{_{w/o\;cold}}$} can help models improve performance as it heuristically removes some noises from user histories. However, a more reasonable and rational way may be to reserve all products and leave the process of cold products to the model designer and researchers. From the performances in \textbf{JDsearch$\rm{_{short}}$} and \textbf{JDsearch$\rm{_{long}}$} dataset, we can find that personalized models usually can achieve high results on users who frequently do online shopping but perform badly while facing cold users. This finding can inspire future research applying different personalization strategies for users with different personalities.

\section{Discussion of Application scenarios}
Besides utilizing it to run experiments and evaluate performances for personalized product search models, the JDsearch dataset can also support research in other areas. First, researchers can simply ignore the query information or remove the interactions under queries in the dataset to use it to assess product recommendation models. In addition, since we record all user behaviors from diverse channels including both search and recommendation, the JDsearch dataset can also support studies in the unified recommendation and search model, which has no publicly available datasets to the best of our knowledge.

\section{CONCLUSION}
\label{sec:conclusion}
In this work, we introduce a new personalized product search JDsearch dataset collected from real user behaviors. Different from the simulated Amazon sub-category dataset, our dataset includes real user queries. Besides, there also exist products belonging to various categories. Cold products and users are also included. These lead the JDsearch dataset closer to the real product search situations. We also investigate the characteristics of the dataset from several perspectives and test existing personalized product search models. These analyses and experiments verify the feasibility of the proposed dataset. This dataset can also support some potential personalization directions including few-shot scenarios, multi-interest modeling, and separate strategies for different types of users.

\section*{Acknowledgments}
Zhicheng Dou is the corresponding author. This work is done during Jiongnan Liu's internship at JD. This work was supported by the National Key R\&D Program of China (2022ZD0120103), National Natural Science Foundation of China (62272467),  Beijing Outstanding Young Scientist Program NO. BJJWZYJH012019100020098,  Public Computing Cloud, Renmin University of China, and Intelligent Social Governance Platform, Major Innovation \& Planning Interdisciplinary Platform for the ``Double-First Class'' Initiative, Renmin University of China. The work was partially done at Beijing Key Laboratory of Big Data Management and Analysis Methods.

\bibliographystyle{ACM-Reference-Format}
\balance
\bibliography{sample-base}

\end{document}